\documentstyle[preprint,aps,version2]{revtex}
\begin{document}
\draft 
\begin{title}
Persistent Currents From the Competition\\
Between Zeeman Coupling and Spin-Orbit Interaction
\end{title}
\author{Tie-Zheng Qian, Ya-Sha Yi and Zhao-Bin Su}
\begin{instit}
Institute of Theoretical Physics, Academia Sinica,\\
P.O. Box 2735, Beijing 100080, 
People's Republic of China 
\end{instit}
\begin{abstract}
Applying the non-adiabatic Aharonov-Anandan phase approach to a
mesoscopic ring with non-interacting many electrons 
in the presence of the spin-orbit interaction,
Zeeman coupling and magnetic flux, 
we show that the time-reversal symmetry breaking due to Zeeman coupling 
is intrinsically
different from that due to magnetic flux. We find that the 
direction of the persistent currents induced by the Zeeman 
coupling changes periodically with the particle number, 
while the magnetic flux
determines the direction of the induced currents by its sign alone.
\end{abstract}

\narrowtext
The persistent currents in multiply connected mesoscopic systems have
attracted much attention in recent years \cite{1}. Conventionally, 
the persistent
currents are produced by applying the magnetic flux to the system and
regarded as one of the famous manifestations of the Aharonov-Bohm (AB)
effect. Through a standard current-current coupling, 
the AB effect only involves the orbital degree of freedom. Since electrons
have spin as well as charge, it is therefore of interest to study if
the spin degree of freedom can play some effective 
and significant role in the persistent current phenomena.

Recently, based on the discovery of the geometric phase \cite{2}, 
many authors have investigated the persistent currents induced by the
geometric phases which originate from the interplay between electrons' orbital
and spin degrees of freedom.   
Such interplay can be maintained by     
external electric and magnetic fields, which lead to spin-orbit (SO) 
interaction and Zeeman coupling respectively. 
Loss {\it et al}. studied the textured ring embedded in inhomogeneous
magnetic field \cite{3}.  
On the other hand,  Meir {\it et al}. showed  
that the SO interaction in one-dimensional rings results in  
an effective magnetic flux \cite{4}.
Mathur and Stone \cite{5} then pointed out that observable phenomena induced
by the SO interaction are the manifestations of the Aharonov-Casher (AC) 
effect \cite{6} in electronic
systems.  Balatsky and Altshuler \cite{7}, Choi \cite{8} and Oh and Ryu \cite{9}
studied the persistent currents produced by the AC effect.

So far the effects of the SO interaction and Zeeman coupling
on the persistent currents have only been discussed separately.
As is well known, the SO interaction and Zeeman coupling are of quite different
time-reversal transformation properties. The SO interaction is time-reversal
invariant and results in the AC flux which induces
the charge current in each single-particle electronic state.
However, at finite temperature the thermal-equilibrium persistent currents
always vanish in systems which possess the time-reversal symmetry (TRS).
On the other hand, the Zeeman coupling breaks the TRS. 
In particular, the finite-temperature persistent current is 
essentially a manifestation
of the breaking of the TRS due to certain external influence. 
We therefore expect that in a many-electron system
the Zeeman coupling can induce the persistent currents 
by competing with the SO interaction
which can not produce the finite-temperature persistent current by itself but 
provides a complete set of current-carrying single-particle states.
This implies that the Zeeman coupling can induce the persistent currents in
a way which is totally independent of its inhomogeneity that results in
the geometric phase \cite{3}.

The purpose of this paper is to reveal the role of the 
TRS in the persistent current phenomena when the spin
degree of freedom is fully taken into account explicitly.
We focus on the low-temperature persistent charge currents in a non-interacting
many-electron ring in the presence of {\it both} the Zeeman coupling 
and SO interaction. As shown by Kravtsov and Zirnbauer \cite{10},
the AB flux can induce the persistent currents by lifting 
the Kramers degeneracy
which in general exists in time-reversal invariant systems with odd
number of electrons. 
In this paper, we demonstrate that in the presence of the SO interaction  
which induces the currents opposite in direction but equal in magnitude
in the two degenerate single-particle states of each Kramers doublet, as a 
result of the simultaneous spin and orbital quantum number
dependence of the energy value and current magnitude for each doublet,
the TRS breaking mechanism due to the Zeeman coupling is intrinsically
different from that due to the AB flux and accordingly 
significant observable effects for the persistent currents exist. 
The paper is organized as follows. First, the exact solution of the 
Schr\"{o}dinger equation for the system 
in presence of the SO interaction, Zeeman coupling and AB flux 
simultaneously is derived in an approach
based on the concepts of spin cyclic evolution and the corresponding 
Aharonov-Anandan (AA) phase \cite{11,12}. Then, with the explicit expression
of the persistent currents derived, the TRS breaking effects of the Zeeman
coupling and AB flux upon the persistent currents 
are studied and compared with each other. 
It is found that the 
direction of the currents induced by the Zeeman 
coupling changes periodically with the particle number, while the AB flux
determines the direction of the induced currents by its sign alone. 
Throughout the discussion, we emphasize that the observability of
the above mentioned symmetry breaking effects can 
only be ensured in systems with a strong SO interaction that can 
produce an AC phase of order unity.
We also point out that the zero-temperature and
low-temperature persistent currents derived for the perfect rings
does not change qualitatively if disordered rings are considered.
As the observed current magnitude in a 
strongly interacting GaAs-AlGaAs ring agrees well with
the theoretical result for a non-interacting system without disorder \cite{13},
we finally have a numerical estimation
for a ring formed by two-dimensional electron gas (2DEG) on a semiconductor
heterostructure, which provides a strong effective electric field
and makes experimental observation possible.

For a 1D ring lying in the
$xy$ plane with its center at the origin,
in the presence of both electric and magnetic fields
${\bf E}=E(\cos\chi_{1}{\bf e_{r}}-\sin\chi_{1}{\bf e_{z}})$
, ${\bf B}=B(\sin\chi_{2}{\bf e_{r}}+\cos\chi_{2}{\bf e_{z}})$
in the cylindrical coordinate system, the 
one-particle Hamiltonian for non-interacting electrons is given by
\begin{equation}
H=  \displaystyle\frac{\hbar^{2}}{2m_{e}a^{2}}
[-i\displaystyle\frac{\partial}{\partial \theta}
+\phi+\alpha 
(\sin\chi_{1}\sigma_{r}+\cos\chi_{1}
\sigma_{z})]^{2} +\displaystyle\frac{\hbar 
\omega_{B}}{2}(\sin\chi_{2}\sigma_{r}+\cos\chi_{2}\sigma_{z}),
\end{equation}
with $\sigma_{r}=\sigma_{x}\cos\theta 
+\sigma_{y}\sin\theta $,
$\alpha =-\frac{eaE}{4m_{e}c^{2}}$ and
$\omega_{B} =-\frac{geB}{2m_{e}c}$, 
where $a$ is the ring radius, $\theta$ is the angular coordinate
and $\phi$ is the enclosed AB flux in unit of flux quantum. 
The exact eigenfunctions $\Psi_{n,\mu}$ and eigenvalues 
$E_{n,\mu}$ of Hamiltonian (1) are obtained as follows
\begin{equation}
\Psi_{n,\mu}(\theta)=
{\rm exp}(in\theta)\tilde{\psi}_{n,\mu}(\theta )/\sqrt{2\pi}; 
\;\;\;\;\;\;
\mu =\pm,
\end{equation}
with 
$\tilde{\psi}_{n,+}(\theta)=\left [ \begin{array}{c}
\cos\frac{\beta_{n}}{2}\\
e^{{\it i}\theta}\sin\frac{\beta_{n}}{2}
\end{array} \right ]$ and  
$\tilde{\psi}_{n,-}(\theta)=\left [ \begin{array}{c}
\sin\frac{\beta_{n}}{2}\\
-e^{{\it i}\theta}\cos\frac{\beta_{n}}{2}
\end{array} \right ]$,
and
\begin{equation}\begin{array}{ll}
 E_{n,\mu}= &\displaystyle\frac{\hbar\omega_{0}}{2}(n+\phi)^{2}
+\displaystyle\frac{\hbar\omega_{0}}{2}(\alpha^{2}-\alpha \cos\chi_{1})
+\displaystyle\frac{\hbar\omega_{n}}{2}(1-\mu\cos\beta_{n})\\&
+\mu\alpha\hbar\omega_{n}\cos(\beta_{n}-\chi_{1})
+\displaystyle\frac{\mu\hbar\omega_{B}}{2}\cos(\beta_{n}-\chi_{2}),
\end{array} 
\end{equation}
where $\omega_{n}=(n+\phi +\frac{1}{2})\omega_{0}$ and
$\beta_{n}$ is given by
\begin{equation}
\tan\beta_{n}= 
\displaystyle\frac{2\alpha\omega_{n}\sin\chi_{1}+\omega_{B}\sin\chi_{2}}
{2\alpha\omega_{n}\cos\chi_{1}+\omega_{B}\cos\chi_{2}
-\omega_{n}}.
\end{equation}
>From the eigenfunctions (3), the persistent charge currents 
of the single-particle states can be expressed as  
\begin{equation}
J_{n,\mu}  =n+\phi-({\Phi_{AA}^{\mu}}+{\Phi_{\rm SO}^{\mu}})/{2\pi}, 
\end{equation}
where $\Phi_{AA}^{\mu}=-\pi (1-\mu\cos
\beta_{n})$ is the geometric AA phase and $\Phi_{\rm SO}^{\mu}
=-2\mu\pi\alpha\cos(\beta_{n}-\chi_{1})$ the dynamical phase contributed by 
the SO interaction \cite{11,12}. 

For $\phi=0$ and $\omega_{B}=0$,
Eqs. (3), (4) and (5) yield
\begin{equation}
E_{n,\mu}^{0}=\displaystyle\frac{\hbar\omega_{0}}{2}
(n-\displaystyle\frac{\Phi_{AC}^{\mu}}{2\pi})^{2},\;\;\;
J_{n,\mu}^{0}=n-\displaystyle\frac{\Phi_{AC}^{\mu}}{2\pi}
\end{equation}
where the AC phase $\Phi_{AC}^{\mu}=-\pi 
(1+2\mu p)$  is the 
sum of the AA phase $\Phi_{AA}^{\mu}$ and dynamical phase $\Phi_{\rm SO}^{\mu}$,
with $p=\sqrt{\alpha^{2}-\alpha\cos\chi_{1}+\frac{1}{4}}$ and 
$\Phi_{AC}^{+}+\Phi_{AC}^{-}=-2\pi$ \cite{12}.
This gives the single-particle Kramers doublet: $(\Psi_{n,\mu}^{0},\;
\Psi_{-n-1,-\mu}^{0})$ 
with the degeneracy $E_{n,\mu}^{0}=E_{-n-1,-\mu}^{0}$ and the current 
relation $J_{n,\mu}^{0}=-J_{-n-1,-\mu}^{0}$, in which $n$ is the 
orbital quantum number while $\mu$ is for spin. 
Due to this single-particle 
Kramers degeneracy, the entire set of energy eigenstates can be divided 
into two subsets $\{ E_{n,\mu};\;\mu=\pm\}$, 
mutually degenerate to each other with one to one  
correspondence. 
For purpose of later application, we divide $p$ into its integer 
part $m$ and fractional part $f$ $(-0.5\le f<0.5)$.
We note that
if $\Phi_{AC}^{\mu}=0$ or $\pi ({\rm mod}\; 2\pi)$, i.e., $f=-0.5$ or $0$, 
then additional degeneracy appears and 
in each branch $\{ E_{n,\mu}^{0}\}$ 
the two single-particle eigenstates with opposite currents
become degenerate. 
Actually, such additional degeneracy is lifted by any weak disorder and
consequently each of the single-particle states carries zero current, 
as illustrated in Fig. 1.
As we will see later, so long as the SO interaction is strong enough to
result in an AC phase of order unity,
a complete set of current-carrying single-particle states resulting from
the fractional $f$ 
is essential to the observability of the low-temperature persistent
currents that the Zeeman coupling or AB flux induces by lifting of the 
Kramers degeneracy. 
Besides the generation of single-particle currents,
the fractional part of the AC flux also plays an 
interesting role in the filling of the Fermi sea of the mesoscopic rings.

For small $\phi$ and $\omega_{B}$, the eigenvalues are of the form
\begin{equation}
E_{n,\mu}=E_{n,\mu}^{0}+\mu\hbar\omega_{B}q/2p+
\hbar\omega_{0}(n+1/2+\mu p)\phi,
\end{equation}   
with $q=\alpha\cos(\chi_{1}-\chi_{2})-\frac{1}{2}\cos\chi_{2}$.
The persistent current $J_{n,\mu}$ also changes from $J_{n,\mu}^{0}$.
But this only leads to high-order correction to the many-electron
thermal-equilibrium persistent currents we will derive.  
Using Eqs. (5), (6) and (7), we can easily
derive the low-temperature charge currents for small $\phi$ and $\omega
_{B}$ in the non-interacting many-electron ring.
For particle number $M$, there are three cases with $M=2N$, $M=4N+1$
and $M=4N+3$ respectively, where $N$ is an arbitrary integer.
We find the charge current for $M=2N$ is always zero when $\phi =0$
and $\omega_{B}=0$, even at zero temperature.
This is actually a result of the Kramers theorem 
which states that the Kramers degeneracy due to the TRS only exists
for odd $M$. Accordingly, the even $M$ ground state at $\phi =0$ and 
$\omega_{B}=0$ 
is a singlet which carries no net current. It can be deduced that
the dependence of the even $M$ zero-temperature currents on the TRS breaking
couplings is smooth and analytical. Kravtsov and Zirnbauer \cite{10}
have shown further that, for odd $M$
electronic systems, the AB flux lifts the Kramers degeneracy and induces 
the persistent currents which vary discontinuously with the flux at zero 
temperature.

For $M=4N+1$ and $4N+3$, if $\omega_{B}<\! <N\omega_{0}$, $\phi <\! <1$,
and $T<\! <N\hbar\omega_{0}/k_{B}$,
we can consider only two single-particle states at and near the Fermi level 
to derive the currents because other states are much higher in energy.
One is the highest occupied state $\Psi_{n_{F},\rho}$
and the other is the lowest unoccupied state $\Psi_{-n_{F}-1,-\rho}$.
These two states form the Kramers doublet
as $\omega_{B}$ and $\phi$ equal to zero.
Denoting their energies as $\epsilon_{\pm}=\epsilon_{0}\pm\epsilon$
and currents as $J_{\pm}=\pm J$, we have the thermal-equilibrium
currents 
\begin{equation}
<J>=-J\tanh(\epsilon /k_{B}T) 
\end{equation}
To justify the above simplification, the energy level spacing
between $E_{n_{F},\rho}$ and $E_{-n_{F}-1,-\rho}$
must be much smaller than the unperturbed spacing
between $E_{n_{F},\rho}^{0}$ (or $E_{-n_{F}-1,-\rho}^{0}$)
and its nearest neighbor in energy in the branch
$\{ E_{n,\rho}^{0}\}$ (or $\{ E_{n,-\rho}\}$).
The level spacing of the latter kind 
is approximately $2|n_{F}|\hbar\omega_{0}\bar{f}$ for large 
$|n_{F}|\approx N$
with $\bar{f}$ the smaller of $|f|$ and $0.5-|f|$,
and it almost vanishes at 
$\Phi_{AC}^{\mu}=0$ and $\pi ( {\rm mod}\;2\pi )$
if the ring is weakly disordered. 
So, in general, besides the generation of single-particle currents,
the strong SO interaction which results in
the AC phase of order unity is also necessary to the validity of 
the above two-state approximation.
We depict the current-carrying characteristic of the single-particle states 
and their level spacing in Fig. 1 to illustrate these two essential
elementary facts which result from the AC phase of order unity and
constitute the basis of the following derivation.

Using Eqs. (6) and (7), for the system with Hamiltonian (1) and odd $M$
non-interacting electrons, we obtain the thermal-equilibrium
persistent currents  
\begin{equation}
<J>_{M}=-J_{F,+}\tanh 
\{\displaystyle\frac{1}{k_{B}T}[\displaystyle\frac{q}{2p}\hbar\omega_{B}+
\hbar\omega_{0}J_{F,+}\phi]\},
\end{equation}
with $J_{F,+}=n_{F}^{+}+\frac{1}{2}+p$,
where for $M=4N+1$, $n_{F}^{+}=N-m$ if $-0.5<f<0$ or $n_{F}^{+}=-N-1-m$ 
if $0<f<0.5$; and
for $M=4N+3$, $n_{F}^{+}=-N-1-m$ if $-0.5<f<0$ or $n_{F}^{+}=N-m$ 
if $0<f<0.5$.
In deriving the two alternate cases of $M$, the filling with respect to 
the splitting of the Kramer doublet plays a crucial role and will be 
explained below.
As $T$ approaches $0^{+}$, the dependences of $<J>_{4N+1}$ and $<J>_{4N+3}$
on $\phi$ and $\omega_{B}$ both tend to step-functions. 

For large $N$, we see $J_{F,+}$ changes its sign from $M=4N+1$ to
$4N+3$, but with its magnitude almost the same.
This leads to $<J>_{4N+1}\approx -<J>_{4N+3}$ for $\phi =0$, 
and $<J>_{4N+1}\approx <J>_{4N+3}$ for $\omega_{B}=0$.
So there exists an interesting and striking difference between
the symmetry breaking effects of the Zeeman coupling and AB flux.
The origin of such difference is that when lifting the Kramers degeneracy,
the Zeeman coupling chooses the ground state with a specific spin orientation 
while the AB flux chooses the
ground state with a specific current direction, as illustrated in Fig. 2. 
This is obviously presented in Eq. (7) in which the energy shift
comprises the two parts. One is of the form $\mu\hbar\omega_{B}$ in which only
spin orientation matters while the other is of the form $J_{n}\phi$ in which only
the sign of the current matters in finding the state with lower energy.
Hence, under the influence of the AB flux, the sign of the 
ground state current never changes with the particle number. 
But for the Zeeman coupling, in stead of the sign of the current,
the spin orientation of the ground state remains the same. 
We can understand why the Zeeman coupling makes the current sign 
change periodically with the particle number as follows. 
Suppose the single-particle state at the Fermi level is $\Psi_{n_{F},\rho}$,
with $\rho= +$ if $q\omega_{B}/p<0$ $($or $\rho=-$ if $q\omega_{B}/p>0)$,
when adding two electrons to the system, first the single-particle state 
$\Psi_{-n_{F}-1,-\rho}$ just above the Fermi level is occupied. 
The second state that becomes occupied
is $\Psi_{m_{F},\rho}$ and $m_{F}$ is determined by the condition that  
$E_{m_{F},\rho}^{0}$ is just above $E_{n_{F},\rho}^{0}$ in the branch
$\{ E_{n,\rho}^{0}\}$.
In short, when adding the electrons two by two, it is effectively
to let the single-particle states  in $\{\Psi_{n,\rho}\}$ occupied one by one 
according to the single-particle
energy levels $\{E_{n,\rho}\}$ which are only different from  
$\{E_{n,\rho}^{0}\}$
by a uniform shift $\rho\hbar\omega_{B}q/p$. This picture is true even if
the ring is disordered. 
It is easy to verify from Eq.(6) that due to the non-integer part of 
$\Phi_{AC}^{\rho}/2\pi$, the electron filling is not fully symmetric with 
respect to the quantum number $n+\rho m$. More explicitly, 
the fraction $1/2+\rho f$ plays an interesting role which 
makes the orbital quantum number $n_F$ of the top electron 
take the values $N-\rho m$ and $-N-1-\rho m$  
for the two alternate cases of odd $M$ respectively.
As a result, for large $|n_{F}|$, 
$m_{F}\cong -n_{F}$. This is to say that 
the ground state current changes its sign periodically
with the particle number 
and such sign dependence can survive at low temperature.

So far our results have been obtained for non-interacting
perfect and disordered rings.
It is worthwhile to further investigate the influence of electron-electron
interaction on the results. 
For the persistent currents in a 
strongly interacting GaAs-AlGaAs ring,
the observed current magnitude is in good agreement
with the theoretical prediction for a non-interacting system without disorder
\cite{13}.
We therefore expect our results are at least qualitatively 
and probably quantitatively related to 
some real systems with very weak disorder.

We make a numerical estimation for a InAs ring \cite{14}. The 
Hamiltonian is of the form
\begin{equation}
H_{\rm InAs}=\displaystyle\frac{1}{2m}({\bf p}-\displaystyle\frac{e{\bf A}}  
{c})^{2}+\hbar\kappa [\sigma \!\!\!\!
\sigma\times{\bf p} ]_{z}-\displaystyle\frac{ge\hbar}{4mc}
\sigma\!\!\!\!\sigma\cdot{\bf B},
\end{equation}
where $m=0.023m_{e}$ is the effective mass, $\hbar^{2}\kappa=6.0\times 10^{-10}
{\rm eVcm}$ is the SO coefficient and $g=15$. 
Here the effective electric field is in the z-direction, hence $\chi_{1}=
\pi /2$. For the loop of radius $a=1 
\mu{\rm m}$, the dimensionless coefficient $\alpha$ is found to be 
$ma\kappa =1.8$ which is large enough to result in an AC phase of order unity
\cite{12}.
The Fermi velocity $v_{F}$ is
approximately $3\times 10^{7} {\rm cms}^{-1}$, corresponding to 
$|n_{F}|\approx 60$ and $I_{F}=ev_{F}/2\pi a\approx 8 {\rm nA}$.
Using the exact solution in Eqs. (2), (3) and (4), we 
numerically compute the charge currents due to the Zeeman coupling
in the non-interacting system with fixed particle number. 
The dependence of
the currents on the magnetic field is depicted in Fig. 3 for various
temperatures, with the zero-temperature limit as a step function.
The striking dependence of the sign or the direction of the charge currents 
on the particle number is also depicted in Fig. 3.
We expect these phenomena can be observed in experiments within the reach of
existing technology.

In conclusion, we have shown that in the presence of strong SO interaction,
the time-reversal symmetry breaking due to the 
Zeeman coupling is intrinsically
different from that due to the AB flux. We have found that the 
direction of the persistent currents induced by the Zeeman 
coupling changes periodically with the particle number, 
while the AB flux
determines the direction of the induced currents by its sign alone. 

T. Z. Qian and Y. S. Yi would like to thank Prof. C. L. Wang for kind help.

\figure
{The current-carrying single-particle states and their level spacing 
caused by the AC phase of order unity. 
$\pm 2\pi\delta=\Phi_{AC}^{\pm}({\rm mod}\;
2\pi)$. The vertical and horizontal arrows indicate the spin orientation and
current direction respectively. 
\label{fig1}}

\figure
{The single-particle energy levels in the presence of SO interaction only
and their shift caused by AB flux and Zeeman coupling. 
\label{fig2}}

\figure
{The charge currents in the InAs ring. 
The dashed and dashed-dotted lines are for $M=4N+1$ and $M=4N+3$ at
$T=10 {\rm mK}$. The solid and dotted
lines are for $M=4N+1$ and $M=4N+3$ at
$T=100 {\rm mK}$.
\label{fig3}}

\end{document}